\documentclass[a4paper]{article}

\usepackage{cite}
\usepackage{color}

  \usepackage{graphicx}
\begin{document}

\title{Silicon Nitride MOMS Oscillator for \\ Room Temperature Quantum Optomechanics}

\author{Enrico~Serra,  
        Bruno~Morana,  
        Antonio Borrielli, 
        Francesco Marin,
        Gregory Pandraud,\\ 
        Antonio Pontin, 
        Giovanni Andrea Prodi, 
        Pasqualina M. Sarro,
        and~Michele~Bonaldi 
\thanks{E.Serra is  with Istituto Nazionale di Fisica Nucleare, TIFPA, 38123 Povo (TN), Italy (e-mail:enrico.serra@tifpa.infn.it); B. Morana, G.Pandraud and P.M. Sarro are with ECTM-EKL, Delft University of Technology, 2628 Delft, The Netherlands (e-mail: b.morana@tudelft.nl; g.pandraud@tudelft.nl;  p.m.sarro@tudelft.nl); A. Borrielli, M. Bonaldi  are with institute of Materials for Electronics and Magnetism, Nanoscience-Trento-FBK Division, 38123 Povo (TN), Italy (e-mail: borrielli@fbk.eu; mbonaldi@fbk.eu); F. Marin and A. Pontin are with Dipartimento di Fisica e Astronomia and LENS, Universit\`a di Firenze, and INFN Sezione di Firenze, Via Sansone 1, 50019 Sesto Fiorentino (FI), Italy and with CNR-INO, L.go Enrico Fermi 6, 50125 Firenze, Italy (email: marin@fi.infn.it; antonio.pontin@gmail.com); G.A. Prodi  is with Dipartimento di Fisica, Universit\`a di Trento, 38123 Povo (TN), Italy (e-mail: giovanniandrea.prodi@unitn.it). The authors gratefully acknowledge the EKL IC process group  for their technological support. Project QuaSeRT has received funding from the QuantERA ERA-NET Cofund in Quantum Technologies implemented within the European Union's Horizon 2020 Programme. The research has been partially supported by INFN (HUMOR project).}
        }

\maketitle

\begin{abstract}
Optomechanical SiN nano-oscillators in high-finesse Fabry-Perot cavities can be used to investigate 
the interaction between mechanical and optical degree of freedom for ultra-sensitive metrology and 
fundamental quantum mechanical studies. 
In this work we present a nano-oscillator made of a high-stress round-shaped SiN membrane with an 
integrated on-chip 3D seismic filter properly designed to reduce mechanical losses. This oscillator works in the 200 kHz - 5 MHz range and features a mechanical quality factor of $Q\simeq10^7$ and a  Q-frequency product in excess of $6.2 \times 10^{12}$ Hz 
at room temperature, fulfilling the minimum requirement for quantum ground-state cooling of the oscillator in an optomechanical cavity. 
The device is obtained by MEMS DRIE bulk micromachining with a two-side silicon 
processing on a Silicon-On-Insulator (SOI) wafer. 
The microfabrication process is quite flexible and additional layers could be deposited over the SiN membrane before the DRIE steps, if required for a sensing application. Therefore, such oscillator is a promising candidate for quantum sensing applications in the context of the emerging field of Quantum Technologies. 
\end{abstract}

%---------------------
\section{Introduction}
%---------------------

{L}{ight} facilitates exploration of quantum phenomena and enables radical 
new technologies. Coupling light with resonant mechanical systems in an optical cavity (Fabry-Perot), 
known as cavity optomechanics \cite{Favero2009, Aspelmeyer2014}, serves as a basis for fundamental studies 
and is a flexible tool for a wide range of scientific and technological goals. We mention for instance 
the sensing of forces or fields at the ultimate limits imposed by quantum mechanics 
\cite{Metcalfe2014, Pontin2018PRA}, and experiments testing the foundation of physics,
i.e., fixing significant constraints for the development of quantum gravity theories \cite{Pikovsky2012, Bawaj2015}. 

Effects related to the quantum fluctuations of the radiation pressure are not easily 
detected because of classical noise sources, like the Brownian thermal noise and spurious coupling of the 
optomechanical system to the environment. As the number of quantum-coherent oscillations in the presence of thermal 
decoherence scales with the Q$\times$f product (mechanical quality factor $\times$ oscillator frequency), this parameter is commonly 
used as a figure of merit for evaluating the oscillator's performance in optomechanics \cite{Aspelmeyer2014}.
For this reason Silicon nitride (SiN) micro- and nano-mechanical membrane resonators have attracted a lot of attention due to their exceptionally high-Q factors \cite{Thompson2008,Tsaturyan2017}. Systems based on a membrane 
oscillator have shown for the first time the mechanical effect of the quantum noise in the light \cite{Purdy2013Science} 
and achieved one of the first observations of pondero-motive light squeezing \cite{Purdy2013PRX}. 
Moreover, high-Q SiN membranes are interesting because they can be used in 
collective optotomechanics,  for the realisation of entangled states between mechanical modes \cite{Nielsen2016}, 
or in systems working in the \emph{anti-squashing} regime of positive feedback for enhanced sideband cooling 
\cite{DavidApril2017}.

In many cases the oscillators consist of a commercial free-standing high-stress silicon nitride (SiN) 
membrane supported by a silicon frame, where mechanical quality factors up to many millions can be in principle 
obtained \cite{Unterreithmeier2010} thanks to the large tensile stress (of the order of GPa). In these systems the mechanical loss is strongly 
dependent on the mounting, especially for the low frequency modes, and values randomly scattered from 10$^4$ 
to some 10$^6$ are usually observed \cite{Wilson2009}. 
Different approaches can be used to isolate the SiN membrane from its support. For instance in \cite{Nielsen2016} a phononic bandgaps 
allowed the observation of ponderomotive squeezing at moderate cryogenic temperatures, while in \cite{Norte2016,Reinhardt2016,Weaver2016} the resonating part  is decoupled by its substrate by high-aspect-ratio SiN trampolines. These achievements motivate the research on customised SiN resonators for optomechanics, using MEMS wafer-scale approaches and with a precise control of the overall dissipation 
mechanisms \cite{Villanueva2014}.

In fact, the mechanical dissipation in an oscillator determines the thermal fluctuation noise, and quantum behavior can emerge only if  the thermal noise force is  lower than noise of quantum origin, such as radiation pressure shot-noise. The Power Spectral Density (PSD) of  the thermal noise force is $S_{Th}= 4 k_B T \frac{m \omega}{Q} $, where $m$ is the mass of the oscillator, $\omega$ its angular frequency,  and $T$ the temperature. The quality factor $Q$ is simply related to the fraction of mechanical energy loss per oscillation cycle, called $\phi$, as: $Q^{-1}=\phi \equiv \frac{\Delta W_t}{2 \pi W_t}$, with $W_t$  the energy stored in the resonant mode and $\Delta W_t$ the energy loss per cycle. As said above, the figure of merit which is the product of mechanical quality factor $Q$ and  frequency $f$ is used as object function in the device optimization process. Indeed, this figure of  merit determines the number of coherent oscillations in the presence of thermal fluctuations, and the minimum requirement for room-temperature quantum optomechanics is $Q \times f > 6.2 \rm \  THz  $ \cite{Aspelmeyer2014}.

We have recently proposed a novel coupled oscillators model for the mechanical losses in a membrane oscillator \cite{Borrielli2016}, considering the mutual interaction between the membrane and the frame in a recoil losses analysis. We were able to design an effective shield for the losses for all mechanical modes also in the low frequency range.  In this work, we present the whole microfabrication process of a MOMS oscillator built following these design rules,  where a circular SiN nanomembrane \cite{Serra2015AIPA} is integrated on-chip with a silicon loss-shield. The shield works as isolation stage and protects the oscillator from the mechanical decoherence induced by the thermal bath. The device is fully functional from room to cryogenic temperatures, and reaches high $\rm f \times Q$  values for all resonant modes in the measured bandwidth [0.2-5] MHz.
The microfabbrication combines silicon double-side deep-RIE etching on a Silicon-On-Insulator (SOI) wafer with controlled wet-etch HF steps 
for removing sacrificial oxides and controlling the release of the nanomembrane. 
Additional layers can be deposited onto the SiN layer before fabrication, for sensing applications or to tailor the membrane reflectivity,  without increasing the process complexity. 

%-------------------------------------
\section{RESONATOR DESIGN}
\begin{figure}[h!]
\centering
\includegraphics[width=8.5cm]{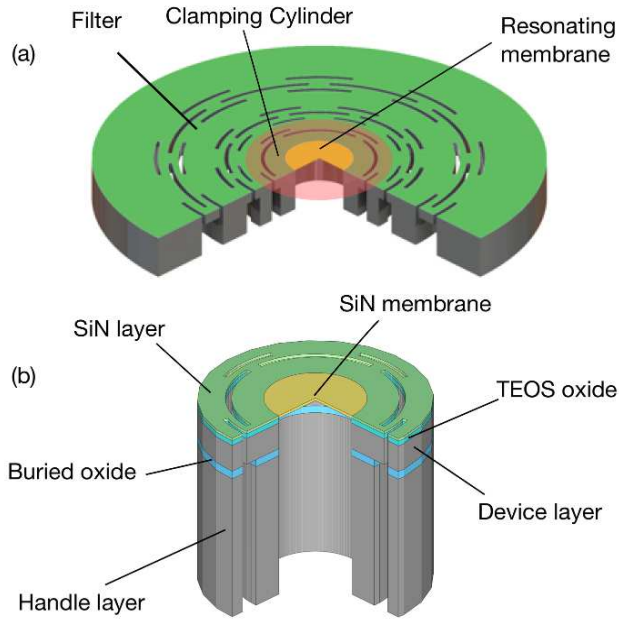}
\caption{(a) View of the SiN circular membrane oscillator. The main oscillator (SiN membrane) 
is in the center and surrounded by the loss-shield structure (filter). (b) Detailed view of the central 
part of the oscillator showing the SOI wafer and thin film layers (not to  scale).}
\label{fig:IEEE_fig1}
\end{figure}

%-------------------------------------
A round-shaped high-stress non-stoichiometric 
SiN membrane (with tensile intrinsic stress of 0.830 GPa), deposited by Low-Pressure Chemical Vapour Deposition (LPCVD), of thickness around 100$\,
$nm is analysed. This membrane interacts by the action of the radiation pressure, with a 1064 nm laser beam. 
To improve the quality factor and to reject vibrational noise from the environment, 
a mechanical filtering stage is integrated on-chip by exploiting double side Deep Reactive-Ion Etching (DRIE). 
A detail view of the device is shown in Fig. \ref{fig:IEEE_fig1}. 
The design of the SiN MOMS oscillator comprises of three main parts: (1) the resonating thin silicon nitride ($\rm SiN$) nanomembrane, (2) the 
loss-shield working as intermediate filter (made of a clamping hollow cylinder and flexural/torsional 
springs), (3) the outer silicon frame that connects the oscillator to the sample holder. 
Fig. \ref{fig:IEEE_fig1}b shows a detailed view of the cross section of the central part of the system, 
highlighting the thin-film layers and the SOI wafer structure.
In the subsections \ref{Total losses}, \ref{Coupled system losses}, \ref{Shielding recoil losses} we focus our discuss on how to minimise the thermal noise effects 
by reducing the mechanical dissipations of the substrate with a loss shield. In subsection \ref{Intrinsic losses} 
we discuss how to limit the detrimental effects on Q-fact due to the oxide layer even at 
liquid-He (4 K) temperatures.

%---------------------------------------------------------
%\subsection{MECHANICAL LOSSES MINIMIZATION}
%---------------------------------------------------------

%-----------------
%-----------------

%------------------------
\subsection{Total losses}
%------------------------
\label{Total losses}

The energy loss in a SiN membrane is dominated by the intrinsic 
loss, that is the imaginary component in the stress-strain relation. This loss sets the ultimate limit $Q_{int}$ for the mode's $Q$-factors, but when the membrane is part of a complex device, different loss channels add up according to this simple relation: 
%---------------
\begin{equation}
Q_{TOT}^{-1}=\phi_{TOT}=\phi_{int}+\phi_{clam}+\phi_{gas}+\phi_{TED}
\label{eq:QTot}
\end{equation}
%---------------
where: 
$\phi_{clam}$ is due  to the clamping loss, $\phi_{gas}$ describes viscous damping arising from gas surrounding the membrane, $\phi_{TED}$ is the thermoelastic damping. We point out that, in a highly stressed membrane oscillator, the total $Q$-factor is also proportional to the internal stress. In fact when the internal stress grows,  the energy dissipated in a time unit remains the same, while the oscillation frequency is much higher. The energy lost per cycle $\phi_{TOT}$, thanks to this \emph{dilution effect}, is reduced and we observe a corresponding improvement of $Q_{TOT}$ \cite{Unterreithmeier2010,Yu2012}. Given that the internal stress cannot be much larger than 1 GPa, the technological efforts are currently directed to reduce contributions from each loss channel.

In our MOMS device we can neglect the gas damping $\phi_{gas}$ because the oscillator operate in a High Vacuum setup, and the thermoelastic damping of the SiN layer is orders of magnitude lower than the other losses 
mechanisms.  
Therefore the most relevant contribution is the clamping loss $\phi_{clam}$, that depends on the detail of the mechanical impedance of the support system. In fact, as observed for instance by Morse \cite{Morse48} for the case of a string supported by a sounding board, the damping depends on the mechanical impedance of the support. Only in the special case of infinitely  rigid and massive
support, the damping of the oscillating part is determined only by its intrinsic dissipation. In case of highly stressed membranes, the typical resonant frequency is around 1 MHz. In this frequency range, the silicon die cannot be considered as a rigid body but its full modal response must be considered in evaluating its admittance, because a support resonating at the same frequency of the membrane can be very effective in absorbing mechanical energy. For this reason we have developed a theoretical model \cite{Borrielli2016} where we evaluate the loss when the  membrane and support are fully-coupled, i.e. allowing for the transfer of energy in both ways. In the next section we evaluate the clamping loss when the membrane and the silicon frame are coupled structures, and we show how the loss can be reduced by using mechanical insulation stages (i.e. a loss-shield structure).

%-----------------
\begin{figure}[h!]
\centering
\includegraphics[width=8.5cm]{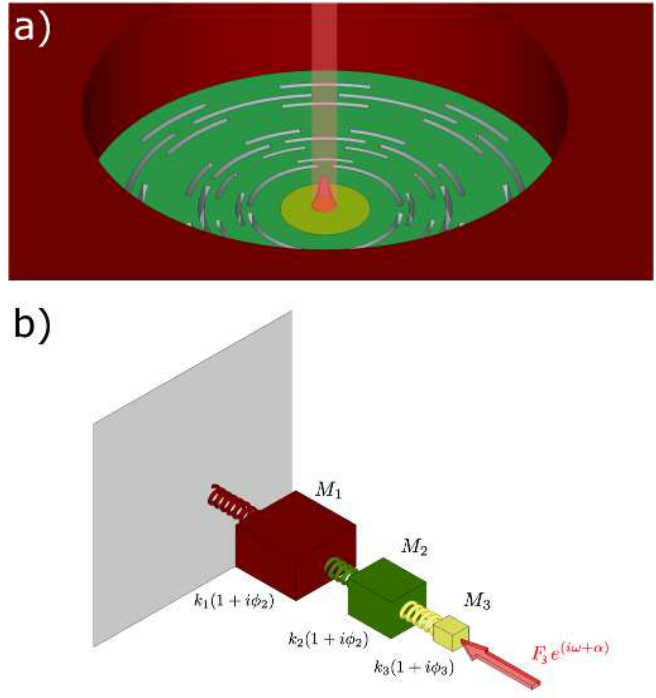}
\caption{(a) Model used for in the FEM analysis for the calculation of the thermal noise and 
the evaluation of the $Q$-factor. (b) Lumped-model for the three resonators (membrane-loss shield-wafer) for the physical 
interpretation of the recoil losses and the evaluation of the thermal noise at the membrane 
surface and of the $Q$-factor }
\label{fig:IEEE_fig23}
\end{figure}
%-----------------
 
%----------------------------------------
\subsection{Coupled system losses}
%----------------------------------------
\label{Coupled system losses}

For a precise evaluation of total loss, we consider two-way energy transfer between membrane and its support. The coupling is originated 
from the recoil forces or torques of the structure surrounding the vibrating SiN membrane. 
The  damping originating from recoil forces has been first evaluated by Saulson \cite{Saulson1990} in the analysis of two cascaded pendula used as mechanical filters in  Gravitational Wave detectors. If we model the membrane of mass $M_m$ and the wafer of mass $M_w$ as two coupled  harmonic oscillator, with $\omega_m$ and $\omega_w$ the resonant frequencies of the the uncoupled parts,  the effective loss angle $\overline{\phi}_m$ of the  membrane  in the coupled system is:
%---------------
\begin{equation}
\overline{\phi}_m(\omega_m)=\phi_m+\phi_w \frac{M_m}{M_w} \frac{\omega_w^2 \omega_m^2}{(\omega_w^2-\omega_m^2)^2}, \ \ \omega_w \ne \omega_m 
\label{eq:eqQmcoupled1}
\end{equation}
%---------------
where $\phi_m, \phi_w$ are the intrinsic loss factor of the membrane and the wafer, respectively. This equation shows that in the case of a massive, rigid support with a small intrinsic dissipation ($\phi_w\simeq 0$), the dissipation of the coupled system is essentially determined by the intrinsic loss in the membrane. 
But if the two parts are  at resonance, i.e. $\omega_w \simeq \omega_m$,  an approximated solution, valid for the case $\phi_m \ll 1$, $10^2<\phi_w^{-1} <10^5$ and $M_w/M_m \gg \frac{1}{\phi_w^2}$, gives:
%---------------
\begin{equation}
 \overline{\phi}_m  \simeq   \phi_m+\frac{1}{\phi_w^2} \frac{M_m^2}{M_w^2}, \ \ \omega_w \simeq \omega_m
\label{eq:eqQmcoupled_res1}
\end{equation}
%---------------
and we see that the loss of the coupled membrane will grow as some mechanical energy leaks toward the supporting wafer.

%---------------------------------------
\subsection{Shielding recoil losses}
%---------------------------------------
\label{Shielding recoil losses}

To overcome the loss due to the coupling between the wafer and the resonating membrane, we introduce a third body with a known dynamical impedance (loss-shield stage) that can effectively decouple the membrane from the support. This intermediate stage is also effective in filtering-out low frequency displacement noise. This approach will make the $Q$-factor independent from the internal resonant modes of the system setup. In contrast, commercial TMAH/KOH wet-etched SiN membranes have a $Q$-factor dependent on the specific clamping implemented in the setup. 

Given that a dissipative system does not admit a rigorous description in terms of normal modes \cite{meirovitch}, we evaluate by a numerical analysis the mechanical susceptibility and the associated loss of the motion of the membrane as seen from the readout port. In our system the optical readout samples the displacement of the surface on a circular area with a gaussian intensity profile centered on the membrane, as shown in Fig. \ref{fig:IEEE_fig23}a. We point out that the results of the analysis are only slightly dependent on the implementation details, and the optimization remains valid if the beam shape is changed or if a radio frequency readout is used.

In Fig. \ref{fig:IEEE_fig23} we show  a CAD image of the 3D real system, a practical implementation of a loss-shield made of flexural and torsional joints and the corresponding three-mode lumped resonating system. In both cases we indicate the application point of the laser beam used for the detection of the displacement. 

As a first step, we calculated the mechanical response of the lumped tree-mode oscillator, driven by a harmonic force. The effective loss (and therefore the quality factor) is evaluated from the imaginary part of the complex susceptibility.  The parameters of the model are then optimized to obtain $Q$-factors of the order of $10^7$. As a second step, we study the 3D model by Finite Element and tune its geometry to match the dynamic admittance  of the optimal three-oscillator model. 

Referring to Fig. \ref{fig:IEEE_fig23}b, in the frequency domain the displacement array is related to the force array by a 3$\times$3 mechanical impedance matrix as:  $ [\tilde{Z}_{ij}(\omega)] \, \tilde\textbf{u}=\tilde\textbf{F}_d$, where $\tilde\textbf{u}=[u_1,u_2,u_3]^T$ is the displacement vector and $\tilde\textbf{F}_d$ is the driving force vector. The tridiagonal matrix $[\tilde{Z}_{ij}(\omega)]$ can be easily obtained as a function of the complex spring constants $\hat{k}_{n}=k_n (1+i \phi_n)$ (with ${k}_n$ the spring constant and  $\phi_n$ the associated loss factor):
% Commands to reduce the matrix
\newcommand{\mysmallarraydecl}{\renewcommand{%
\IEEEeqnarraymathstyle}{\scriptscriptstyle}%
\renewcommand{\IEEEeqnarraytextstyle}{\scriptsize}%
\renewcommand{\baselinestretch}{1.1}%
\settowidth{\normalbaselineskip}{\scriptsize
\hspace{\baselinestretch\baselineskip}}%
\setlength{\baselineskip}{\normalbaselineskip}%
\setlength{\jot}{200\normalbaselineskip}%
\setlength{\arraycolsep}{2pt}}

%--------------
\begin{equation}
[\tilde{Z}_{ij}(\omega)] =\left(\begin{array}{ccc}
-M_{1} \omega^2+\hat{k}_{1} & -\hat{k}_{1}  & 0  \\
-\hat{k}_{1} & -M_{2} \omega^2+ \hat{k}_{1}+ \hat{k}_{2} & -\hat{k}_{2} \\
 0 &-\hat{k}_{2}& -M_{3} \omega^2+ \hat{k}_{2}+ \hat{k}_{3}
\end{array}\right)
\end{equation}
%--------------
Here $(M_{1}, \hat{k}_1)$ are the effective mass and the complex spring 
constant of the support (wafer + OFHC copper support), $(M_{2}, \hat{k}_2)$ are the effective mass and the complex spring 
constant of the loss-shield structure and $(M_{3}, \hat{k}_3)$ are the effective mass and the complex  spring constant 
of the membrane. 
 The formal solution of the dynamical equation is $ \tilde\textbf{u} = [\tilde Y_{ij}(\omega)]\, \tilde\textbf{F}_d$, where $[\tilde{Y}_{ij}(\omega)]=[\tilde{Z}_{ij}(\omega)]^{-1}$ is the admittance matrix of the system.
 This equation could be solved to evaluate the resonant frequencies and the complex normal modes of the system, but on a practical level we need the response of the membrane $M_{3}$ to a single external force  $F_3$, as the membrane's position is the only measured variable in the system. In agreement with the fluctuation-dissipation theorem, this is equivalent to the study of the thermal noise power spectrum of the oscillator, as seen from the readout port \cite{Levin98}: 
 \begin{equation}
 S(\omega)= - \frac{4 k_B T}{\omega}\Im{\left\{ [\tilde Y_{ij}(\omega)]\, [0, 0, \tilde{F}_{3}]^T\right\}}.
\label{eq:eqQmcoupled_res2}
\end{equation}
This curve has a peak at the resonant frequency of the membrane oscillator, and the full width at half maximum of this peak gives a measurement of the quality factor. We point out that this approach estimates the quality factor measured by the dynamical response  of the ``coupled membrane'' (i.e. the membrane coupled to its support system), that is the experimentally accessible quantity.  Given that the membrane is only weakly coupled with its support, its admittance will remain similar in shape to the uncoupled oscillator case: 
%--------------
\begin{equation}
\Im(\tilde{Y}^{c}(\omega))\simeq -\frac{1}{\bar{M}_m}\frac{\omega_m^2 \bar{\phi}_m}{(\omega^2-\bar{\omega}^2_m)^2+\bar{\phi}_m^4 \bar{\omega}^4_m}.
\end{equation}
%--------------
with some proper values of effective mass $\bar{M}_m$,  effective frequency $\bar{\omega}_m$ and effective loss $\bar{\phi}_m$. These effective values depend on the geometrical parameters of the loss shield and wafer, $(M_{1}, \hat{k}_1)$ and $(M_{2}, \hat{k}_2)$. Numerical evaluation show that the corrections to the membrane mass and resonant frequency are negligible ($\bar{M}_m\simeq M_3$ and $\bar{\omega}_m\simeq \omega_3$), while the quality factor $\bar{Q}_m=\phi_m^{-1}$ can become smaller than the uncoupled $Q_3$ if the frequency $\omega_1$ of the wafer approaches $\omega_m$. In this case the loss shield turns out to be very effective in protecting the quality factor of the membrane from excessive reduction.

The optimal parameters of the loss shield, shown in Table \ref{tab:IEEE_table_1} have been found with a numerical approach based on the model of Fig. \ref{fig:IEEE_fig23}b \cite{Borrielli2016}. The actual geometry consistent with these parameters has been developed with the help of FEM modelling.

%----------------
\begin{table}[!t]
\renewcommand{\arraystretch}{1.3}
\caption{Parameters of the uncoupled oscillators used in the 3-oscillator model.}
\label{tab:IEEE_table_1}
\centering
\begin{tabular}{|c|c|}
\hline
& 3-oscillator model\\
\hline
$ M_1$ & $\rm 5 \times 10^{-5}\, Kg$  \\
$\omega_1/2 \pi $& $\rm 227 \, kHz$  \\
$ k_1=M_1 \omega_1^2 $ & $\rm 1.01 \times 10^8\, N/m  $\\
$ \phi_1$ & $\rm 10^{-3} $ \\
$ M_2 $ & $\rm 1 \times 10^{-5}\, Kg$  \\
$ \omega_2/2 \pi $ & $\rm 30 \,kHz$   \\
$ k_2=M_2 \omega_2^2$ & $ \rm 3.55 \times 10^5 \,N/m $ \\
$ \phi_2$ & $10^{-3}  $ \\
$ M_3 $ & $\rm 1.5 \times 10^{-10} \,Kg$  \\
$ \omega_3/2 \pi $ & $\rm 230  \,kHz$   \\
$ k_3=M_3 \omega_3^2 $ & $ \rm 3.13 \times 10^2 \,N/m$   \\
$ \phi_3$ & $10^{-7}   $\\
\hline
\end{tabular}
\end{table}

%----------------------------
\subsection{Intrinsic losses}
%----------------------------
\label{Intrinsic losses}

A critical issue concerning the design of MOMS oscillator is the thin oxide layer that is used 
as etch-stop layer for the Deep-RIE on the back side. The thin TEOS $\rm SiO_2$ layer is shown in 
the design concept of the oscillator Figure \ref{fig:IEEE_fig1}. This layer is completely removed from the membrane in a release step following the Deep-Rie etching of the substrate. This layer, if not removed properly at the membrane's edge, can originate an additional loss depending on its thickness. In particular, the dissipation behaviours of this layer can be critical 
at cryogenic temperatures, commonly required in quantum optomechanics, as $\rm SiO_2$  has a dissipation peak at 50 K \cite{LiuMRS2007} that affects its behaviour down to Liquid Helium  temperatures.

%----------------
\begin{table}[!t]
\renewcommand{\arraystretch}{1.3}
\caption{Material parameters used for evaluating dissipation of the $\rm SiO_2$ layer at 
low and room temperatures (CL - clamping limit - TL thermoelastic limit)}
\centering
\begin{tabular}{|c|c||c|}
\hline
& Low temperature  & Room temperature \\
& ($\rm at \ LHe$) & ($\rm at \ RT$)\\
\hline 
\hline
$E_{\rm SiO_2}\,$(GPa)     & 77  &  77  \\
$ E_{\rm SiN}\,$ (GPa) &  260  & 260 \\
$\rm \rho_{SiO_2} $ $\rm (Kg/m^3)$  & 2800  &  2800   \\
$\rm \rho_{SiN} $ $\rm (Kg/m^3)$  & 2100 & 2100\\
$\rm \sigma_0$ $\rm (MPa)$ & 800 &  800 \\
$\rm \phi_{SiO_2}$ & $\rm 1 \times 10^{-3}$ & $\rm 5 \times 10^{-4}$ \\
$\rm \phi_{SiN}$ & $\rm 2 \times 10^{-5}$ & $\rm 2 \times 10^{-5}$ \\
$\rm \phi_{Si}$ & $\rm 1 \times 10^{-6}$  &  $\rm 3 \times 10^{-4}$ (TL) \\
$\rm \phi_{wafer}$ & $\rm 1 \times 10^{-3} (CL)$  &  $\rm 1 \times 10^{-3} (CL)$  \\
\hline
\end{tabular}
\label{tab:IEEE_table_2}
\end{table}
%----------------

The intrinsic quality factor ($Q_{\rm int}=\phi_{\rm int}^{-1}$) of a free-standing membrane dominated 
by intrinsic stress $\sigma_0$ can be calculated, for a square-shaped membrane, with the approach described in \cite{Yu2012} 
: 
%---------------
\begin{equation}
Q_{\rm int}(n_1,n_2,\sigma_0)=\frac{1}{\lambda} \frac{E^{\rm Re}}{E^{\rm Im}} (1 + \lambda \frac{n_1^2+n_2^2}{4})^{-1}
\label{eq:Qkn}
\end{equation}
%---------------
where $ (n_1,n_2)$ are the mode indexes, $\lambda=\sqrt{2 F/\sigma_0 h l^2}$ 
with $F$ the flexural rigidity, $ h, l$ the thickness and the side length of the square-shaped SiN layer. $E^{\rm Re}$ and $ E^{\rm Im}$ are the real and imaginary parts of the complex Young's modulus, that are related to the loss factor according to the following relation: $ \hat{E}=E^{\rm Re}+ i E^{\rm Im}\equiv E(1+i \phi)$.
We take \ref{eq:Qkn}  as an estimate for the quality factor in the case of a circular membrane with similar frequency. 

In \ref{eq:Qkn}, the physical meaning of $\lambda$ is the ratio between the bending energy 
to the elongation energy for the fundamental mode (in our case $\lambda=1.08 \times 10^{-3}$). The loss at the edge is described by the term independent from the mode indexes $(n_1,n_2)$, that is:
%---------------
\begin{equation}
Q_{\rm edge}(\sigma_0)=\frac{1}{\lambda} \frac{E^{\rm Re}}{E^{\rm Im}}
\label{eq:Qedge}
\end{equation}
%---------------
 
To evaluate the lower limit of the quality factor due to the TEOS oxide layer, we estimate the components of the complex Young modulus by considering the effective thin film layer: 
%---------------
\begin{eqnarray}
 &&E^{\rm Re}=E_{\rm SiN} f_{v_{\rm SiN}}+E_{\rm SiO_2} f_{v_{\rm SiO_2}}\nonumber \\
 && =\frac{E_{\rm SiN} h_{\rm SiN}+E_{\rm SiO_2} h_{\rm SiO_2}}{h_{\rm SiN}+h_{\rm SiO_2}} \\
 &&E^{Im}=E_{ \rm SiN} \phi_{\rm SiN} f_{v_{\rm SiN}}+E_{\rm SiO_2} \phi_{\rm SiO_2} f_{v_{\rm SiO_2}} \nonumber \\
 && =\frac{E_{\rm SiN}\phi_{\rm SiN} h_{\rm SiN}+E_{\rm SiO_2}\phi_{\rm SiO_2} h_{\rm SiO_2}}{h_{\rm SiN}+h_{\rm SiO_2}}
\label{eq:E1E2}
\end{eqnarray}
%---------------
where $E^{\rm Re}$ and $E^{\rm Im}$ are the effective real and imaginary Young modulus, $ \rm f_{v_{SiN}},f_{v_{SiO_2}}$ 
are the volume fractions while $E_{\rm SiN},E_{\rm  SiO_2}$ are the Young modulus and the loss angle 
$\phi_{\rm SiN}, \phi_{\rm SiO_2}$, as reported in Table \ref{tab:IEEE_table_2}. 

\begin{figure}[htbp]
\begin{center}
\includegraphics[width=80mm]{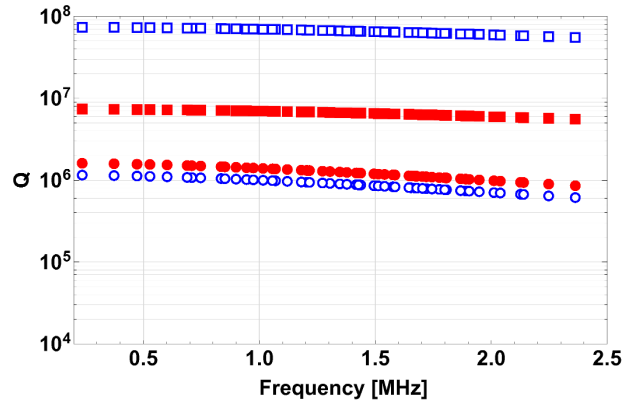}
\caption{Results of FEM simulations. Q-factor at room temperature (300 K): (red-filled squares) 100 nm thick  SiN
membrane; (red-filled circles) bilayer $\rm SiN/SiO_2$ (100/290) nm. Q-factor at Liquid Helium 
(4 K): (blue squares) 100 nm thick $\rm SiN$ membrane; (blue circles) bilayer $\rm SiN/SiO_2$ (100/290) nm.}
\label{fig:IEEE_fig_4}
\end{center}
\end{figure}

These data are used to simulate the total membrane loss by FEM, showing that for a system without oxide the Q-factor can potentially reach $\rm 5 \times 10^7$ at 4 K, while a reduction of the overall Q-factor is expected in the bilayer sistem $\rm SiN/SiO_2$ (100/290) nm. At room temperature the extra dissipation due to the $\rm SiO_2$ layer is negligible. Results at 4 K and 300 K are reported in Figure \ref{fig:IEEE_fig_4}. 

%-------------------
%-------------------

%-------------------------
\section{MICROFABRICATION}
%-------------------------

Resonators were fabricated exploiting MEMS bulk-micromachining by Deep-Reaction Ion Etching
(DRIE) and through wafer two-side processing, a process which has already demonstrated the capability of producing low-loss micro-mechanical systems \cite{SerraAPL2012,BorrielliPRApplied2015}.
%------------------
\begin{figure}[ht!]
\centering
\includegraphics[width=3.5in]{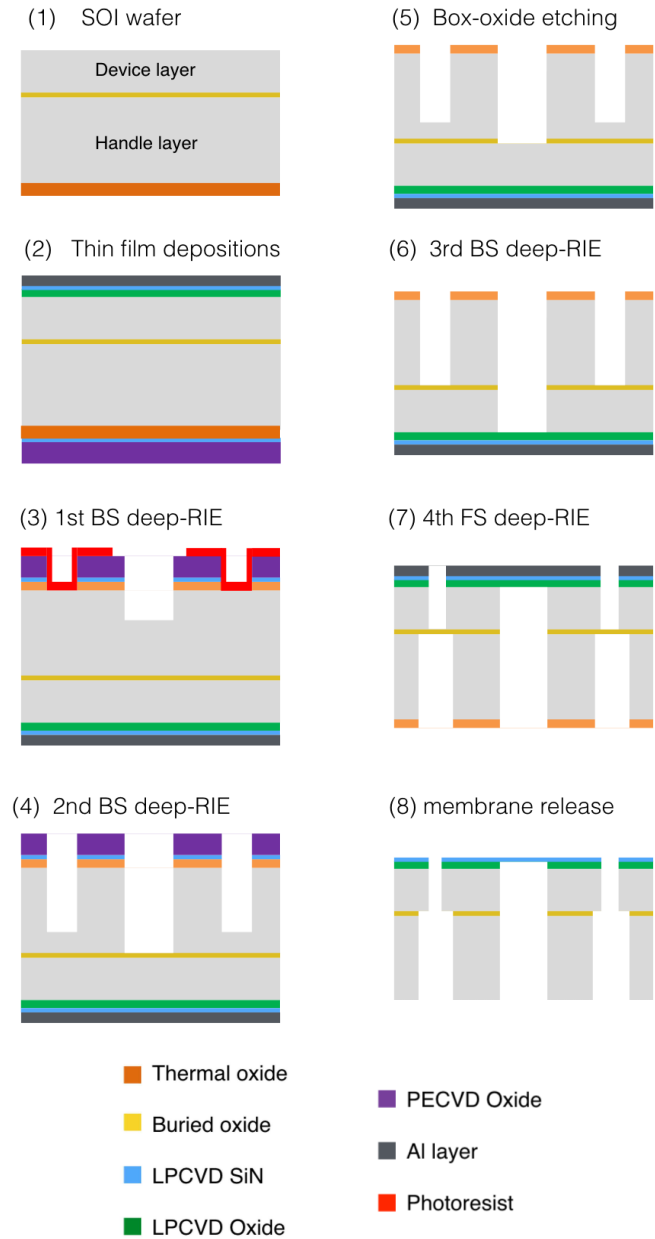}
\caption{(1) SOI wafer; (2) thin film depositions: LPCVD SiN, PECVD $\rm SiO_2$, Al;
(3) $\rm 1^{st}$ DRIE etching on the back-side (BS) with a resist mask with a nested oxide mask; (4) 
photoresist removal and $\rm 2^{nd}$ back-side (BS) etching using the nested oxide mask; (5) $\rm HF$ wet 
etching of the buried oxide; (6) $\rm 3^{nd}$ back-side (BS) etching; (7) front-side (FS) resist AZ9260 and $\rm Al$ 
patterning followed by $\rm 4^{th}$front-side (FS) DRIE etching; (8) membrane release: $\rm Al$ stripping 
by PES chemical bath and oxide HF wet etching.}
\label{fig:IEEE_fig5}
\end{figure} 
%-----------------
Resonators were fabricated from 4-inches  Silicon-On-Insulator (SOI) Floating Zone wafer 
(Icemos Technology Ltd.) of total thickness $\rm 1000 \pm 5 \ \mu m$ (device layer 250 $\mu$m), orientation   $ <100>$ with  resistivity 
$\rm{>1\,k\Omega\,cm}$ (Fig. \ref{fig:IEEE_fig5} step 1). We patterned the handle and device layers by multiple DRIE 
etching steps, using the $\rm{2 \ \mu m}$ buried oxide as etch stop layer.

Firstly, on the device layer we deposited a multilayer thin-film stack composed of a low pressure chemical 
vapor deposition (LPCVD) tetraethyl orthosilicate (TEOS) oxide film with thickness  $\rm 290 \,nm$ (on the bottom), 
the 100-nm-thick LPCVD SiN membrane (in the middle) and  a $\rm 1 \,\mu m$ thick RF-sputtered pure $\rm Al$ layer 
(on the top) as it is shown in Fig. \ref{fig:IEEE_fig5} step 2.
The TEOS oxide works as etch stop layer during DRIE etching steps while the $\rm Al$
layer is used as front side mask and as protection layer for the wet/dry etching processing phases.
The SiN membrane is deposited at about $800^{\circ}{\rm C}$ by tuning the stress to about $0.8 \rm \ GPa$ 
(measured on a Si wafer by curvature methods TENCOR Flexus FLX-2908). Before RF-sputtering of the $\rm Al$ layer, a 
plasma enhanced chemical vapor deposition (PECVD) oxide mask of thickness $\rm 6 \ \mu m$ was deposited 
on the back side at $\rm 400 \ ^{\circ}{C} $ (Fig. \ref{fig:IEEE_fig5} step 2). Silicon etching has been realized by means 
of DRIE two pulse BOSCH process in Omega i2L Rapier at $20^{\ \circ}{\rm C}$. 
A multi-level etching processing was performed to obtain  the membrane's hole and the loss-shield masses,
using resist AZ9260 and the nested PECVD oxide mask. With reference to Fig. \ref{fig:IEEE_fig5} step 3-4 
we first etch $\rm{200 \ \mu m}$ and after stripping the resist we performed a second etching step of $\rm{550 \ \mu m}$ 
landing on the buried-oxide.
%
%-----------------
\begin{figure}[h!]
\centering
\includegraphics[width=8.0cm]{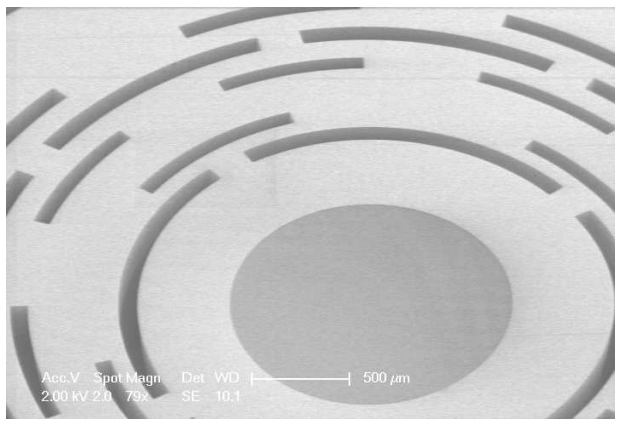}
\caption{SEM image of the front side of the circular SiN membrane 
oscillator with the loss-shield.}
\label{fig:IEEE_fig6}
\end{figure}
%-----------------
%-----------------
\begin{figure}[h!]
\centering
\includegraphics[width=8.0cm]{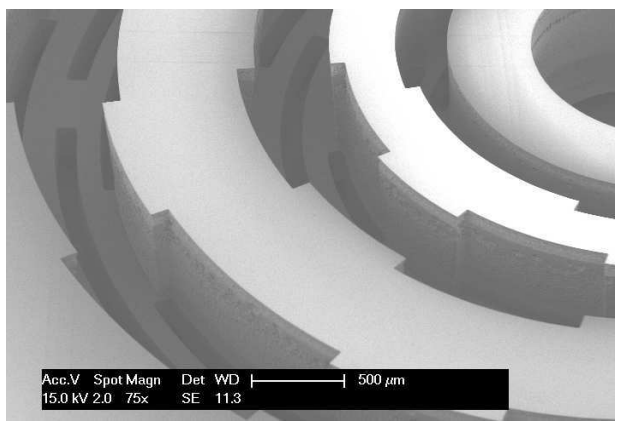}
\caption{SEM image of the back side of the membrane oscillator showing the 
loss-shield masses.}
\label{fig:IEEE_fig9}
\end{figure}
%-----------------

%-----------------
\begin{figure}[h!]
\centering
\includegraphics[width=8.0cm]{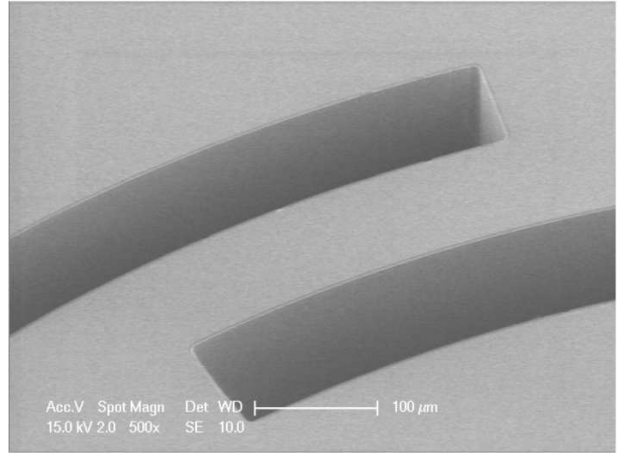}
\caption{SEM image of the front side showing the detail of 
the torsional-flexural joints.}
\label{fig:IEEE_fig7}
\end{figure}
%-----------------
%-----------------
\begin{figure}[h!]
\centering
\includegraphics[width=8.0cm]{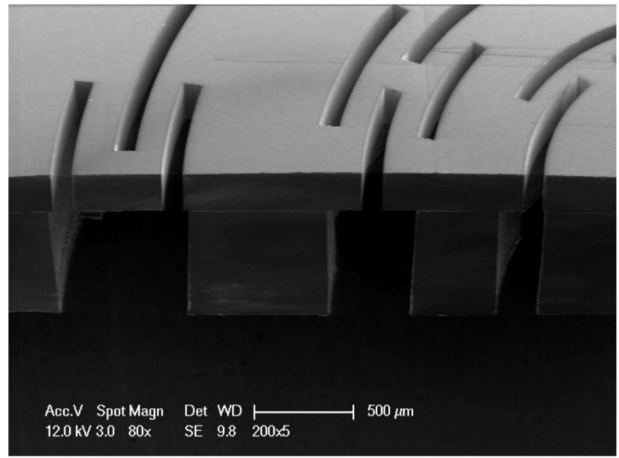}
\caption{SEM image of the cross-section of the device showing torsional-flexural joints 
of the shield-loss stage.}
\label{fig:IEEE_fig_8}
\end{figure}
%-----------------

The estimated average etching rates were about $\rm 1.38\,\mu$m/cycle for Si and $\rm 5 \,$nm/cycle for the PECVD oxide. The gas-flow rate and number of cycles were optimised to control the aspect ratio dependant etching and notching. Afterwards, we removed the buried oxide by means of an HF-based solution (Fig. \ref{fig:IEEE_fig5} step 5) that fully etches the PECVD oxide mask. The SiN layer on the back side works as etch stop layer for HF and protects the thermal oxide underneath. To release the thin-film stack we performed the third DRIE etching step ($\rm{250 \ \mu m}$) exploiting the $\rm 2 \ \mu m$ 
thermal oxide as mask (Fig. \ref{fig:IEEE_fig5} step 6). We divided the etching step in two sub-steps in order 
to prevent membranes failure. We saw re-entrant sidewalls due to aspect ratio dependant etching rates.
By a last DRIE etching step  (Fig. \ref{fig:IEEE_fig5} step 7) we obtained the front-side suspending structure in the $\rm 250 \,\mu m$ device layer; in this case we  used a $\rm 8 \ \mu m$ photoresists AZ9260 (to avoid micro masking of the Al layer). 
Front and back side of the whole oscillator structure are shown in Fig. \ref{fig:IEEE_fig6} and Fig. \ref{fig:IEEE_fig9} respectively, detailed views of the torsional-flexural joints are shown in Fig. \ref{fig:IEEE_fig7} and Fig. \ref{fig:IEEE_fig_8}. 

\begin{figure}[h!]
\centering
\includegraphics[width=8cm]{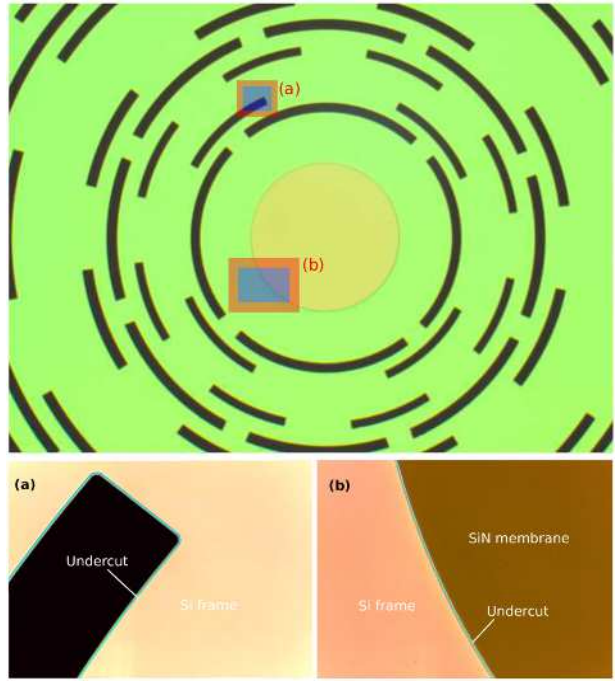}
\caption{(Top) Optical image of the front side of the SiN circular membrane with 
its integrated loss-shield. 
(Bottom) Detailed view of the (a) and (b) areas showing the undercut of $\rm SiO_2$ by HF wet etching for
the membrane realease.}
\label{fig:IEEE_fig_10_11}
\end{figure}

The last step is the SiN membrane release. First we stripped the Al layer using a PES solution, then by means of an HF-based solution we etched the TEOS oxide (Fig. \ref{fig:IEEE_fig5} step 8). 
A HF over-etch is needed in this phase in order to completely release the SiN layer, as any remnant TEOS oxide layer could spoil 
the mechanical Q-factor of the membrane. The appearance of the undercut as it is shown in Fig. \ref{fig:IEEE_fig_10_11} 
(b) represents a good way to set a proper over-etching time. The process ends with dicing and cleaning from resist residues 
by oxygen plasma (TEPLA) and by $\rm HNO_3$ solution $99\%$.

%-----------------
%-----------------

%--------------------------------------
\section{EXPERIMENTAL CHARACTERISATION}
%--------------------------------------

\begin{figure}[h!]
\centering
\includegraphics[width=9cm]{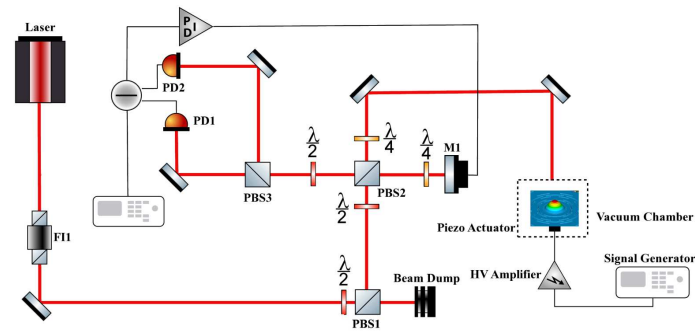}
\caption{Optical scheme of the Michelson 
interferometer apparatus for the measurement 
of the mechanical Q-factor of the modes of the SiN membrane. The main beam is split in PBS2 (polarizing 
beam-splitter) and then recombines in PBS2-PBS3  after reflection from mirror $M_1$ (reference arm) and the SiN membrane 
(sensing arm).}
\label{fig:IEEE_fig_12}
\end{figure}

\begin{figure}[h!]
\centering
\includegraphics[width=8.8cm]{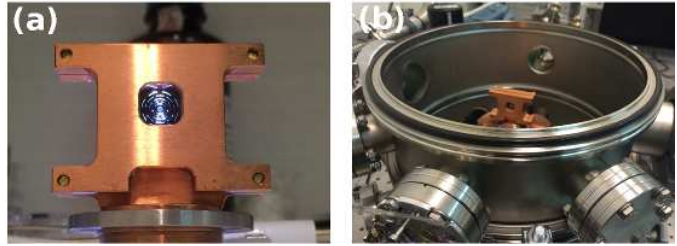}
\caption{(a) Clamping frame made of OFHC copper with the oscillator 
housed in the middle; (b) the system is mounted on top of a seismic filter and housed inside a 
vacuum chamber.}
\label{IEEE_fig_13}
\end{figure}

The experimental set-up used for the measurement of the quality factor is shown in Fig. \ref{fig:IEEE_fig_12}. 
A beam of $\rm 3 \ mW$  reflected by the PBS1 from our light source, a Nd:YAG at $\rm 1064 \ nm$, is aligned 
in a Michelson interferometer followed by a balanced homodyne detection.
In details, a polarizing beam-splitter (PBS2) divides the beam into two parts, orthogonally polarized, 
forming the Michelson interferometer arms. At the end of the first one (reference arm) an 
electromagnetically-driven mirror M1 is used for phase-locking the interferometer in the condition 
of maximum displacement sensitivity. A double pass through a quarter-wave plate rotates by 90$^{\circ}$ the 
polarization of this beam, which is then transmitted by PBS2. The second beam (sensing arm) is focused on 
the membrane oscillator firmly fixed inside the vacuum chamber, and after reflection and double pass through 
the quarter-wave plate is reflected by the PBS2, where it overlaps with the reference beam reflected by M1.
%-----------------
%-----------------
The overlapped beams are then monitored by a homodyne detection, consisting of a half-wave plate, rotating 
the polarizations by 45$^{\circ}$, and a polarizing beamsplitter (PBS3) that divides the radiation into 
two equal parts sent to the photodiodes PD1 and PD2, whose outputs are subtracted. The signal obtained is a 
null-average, sinusoidal function of the path difference in the interferometer. Such a scheme is barely sensitive 
to laser power fluctuations. The difference signal is used as error signal in the locking servo-loop (the locking 
bandwidth is about 1 kHz) and also sent to the acquisition instruments for sensing the displacement of the membrane.
The quality factor of the resonant modes are evaluated by the measurement of the free decay time after resonant excitation with a piezoelectric crystal. When the drive signal is removed, the mechanical vibration follows an exponentially damped decay whose envelope amplitude varies according to: $u(t)=u_0 \exp(-t/\tau_{n_0 n_1})$ where $\tau_{nm}=Q/\pi \nu_{n_0 n_1}$ is the decay time of the mode and $\nu_{n_0 n_1}$ is the resonance frequency of mode with indexes $(n_0,n_1)$, as explained in the next Section. In Fig. \ref{IEEE_fig_13}(a) we show the customized clamping systems for a $\rm 14 \times 14\,   mm^2$ chip and the vacuum chamber Fig. \ref{IEEE_fig_13}(b). The clamping system is isolated from vibrations with a spring-mass system linked to the optical table from the bottom and the pressure in the vacuum chamber is about $\rm 1 \times 10^{-6} \ mbar$ to prevent gas damping effects.
%-----------------

%-------------------------------
\section{RESULTS AND DISCUSSION}
%-------------------------------

\begin{figure}[h!]
\centering
\includegraphics[width=8.5cm]{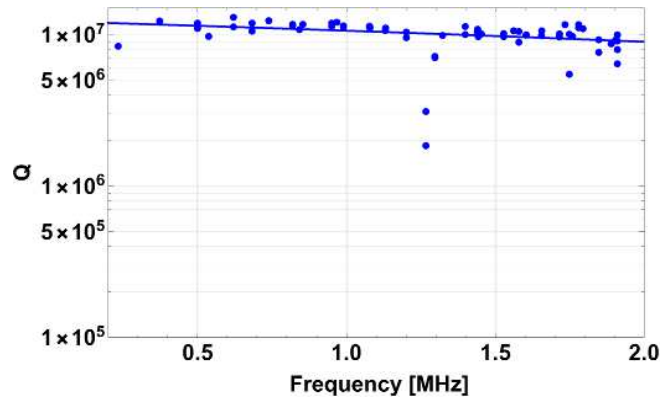}
\caption{The Q-factor at room temperature of  the resonant modes of 
a refernce device  (wf1726-14). We show as a continuous line the least-square regression line of these data, that is in good agreement with the results expected for a square-shaped membrane of the same size \cite{Yu2012}.}
\label{fig:IEEE_fig_18}
\end{figure}

In Fig. \ref{fig:IEEE_fig_18} we report the quality factor of all of the frequency modes (up to 2 MHz) of a reference device (wf1726-14).  In comparison with commercial devices, measured for instance in \cite{Yu2012}, these results highlight the effectiveness of the integrated loss-shield, as almost all modes achieve a high $Q$ value, in good agreement with what is expected by (\ref{eq:Qkn})  for square-shaped 
membranes.  		
 
%----------------
%----------------

\begin{figure}[h!]
\centering
\includegraphics[width=8.8cm]{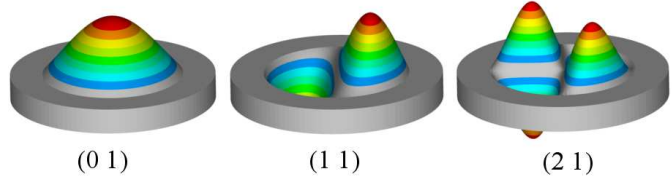}
\caption{Modal shapes the circular membrane with mode indexes $(n_0,n_1)$. The color scale, from
light gray to red, is proportional to the absolute displacement from the equilibrium position.}
\label{modi_v01}
\end{figure}

In general, microfabricated systems should ensure good reproducibility of the operating parameters. To evaluate the robustness of our microfabrication process we compare the $Q$-factor measurements for five devices obtained in five different runs. Relevant 
geometrical parameters related to the thin film oxide and nitride layers are reported in Table 
\ref{tab:IEEE_table_3}. In Table \ref{tab:IEEE_table_4} we report the intrinsic stress $\sigma_0$,  the measured diameter $\rm D$ and the $Q$-factor measurements for three vibrational modes. From these data we estimate the sensitivity of the $Q$-factor to the  variations of the intrinsic stress $\rm \Delta \sigma_0$ and undercut $\Delta U_{ox}$. 
The analysis refer to the three low-frequency oscillations with mode indexes $(n_0,n_1)=(01)/(11)/(21)$, with modal shape shown in Figure \ref{modi_v01}. We point out that for these low frequencies modes a high quality factor can be obtained only thanks to the  loss-shield stage.

\begin{figure}[h!]
\centering
\includegraphics[width=80mm]{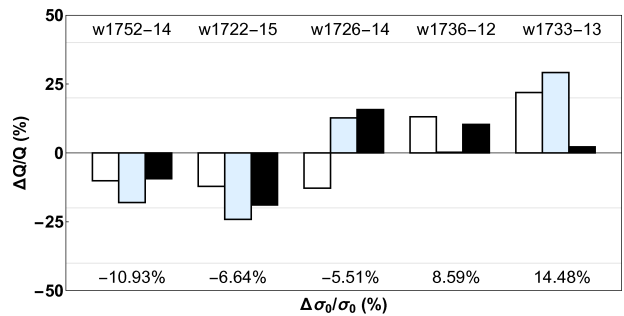}
\caption{Sensitivity of the relative variations of the $Q$-factor, $\Delta Q/Q$, with respect to the relative variation of the stress $\Delta \sigma_0/\sigma_0$ for 
the first three modes:  (white) mode indexes (01);  (light-blue) mode indexes (11); (black) mode indexes (21). The reference values are the average of the measurements over the 5 devices: $753\,$MPa for the stress and $(9.6\times 10^6, 11.06\times 10^6, 9.60\times 10^6)$ for quality factors of the three modes.}
\label{IEEE_fig_14}
\end{figure}

Stress variation are originated by pressure fluctuations and small variations of the  
gas ratio $\rm NH_3/SiH_4 $ during the LPCVD deposition. 
A precise estimate of the intrinsic stress $\sigma_0$ of the membrane, reported in Table \ref{tab:IEEE_table_4}, was derived from the interferometric 
measurement of the frequency $\nu_0$ (Fig. \ref{fig:IEEE_fig_12}). In fact  
the stress and the resonant frequency are connected by the relation: $\nu_{n_0,n_1}=\nu_0 \alpha_{n_0,n_1}$, where $\alpha_{n_0,n_1}$ is the $n_1$-th 
root of the Bessel polynomial of order $ n_0$ and  $\nu_0=\frac{1}{\pi D_0} \sqrt{\frac{\sigma_0}{\rho}}$. Here  
$\rm \rho=2800 \ Kg/m^3$ is the density of the SiN film, and the diameter $\rm D_0$ was directly measured 
on the device by Leica DM750M optical microscope with magnification of 50$\times$. 
According to \ref{eq:Qkn}, the mechanical  $Q$-factor in a square membrane is directly related to the intrinsic stress $\rm \sigma_0$ by the dilution effect of the mechanical losses in SiN films. For this reason it is not surprising that higher quality factors are generally observed when the stress is higher. In Fig. \ref{IEEE_fig_14} we show the relative variation of $\Delta Q/Q$ as a function of the relative variation of stress, taking as reference value the average of the measured stress over the five different devices.

%-----------------
%-----------------

%-----------------
\begin{figure}[h!]
\centering
\includegraphics[width=80mm]{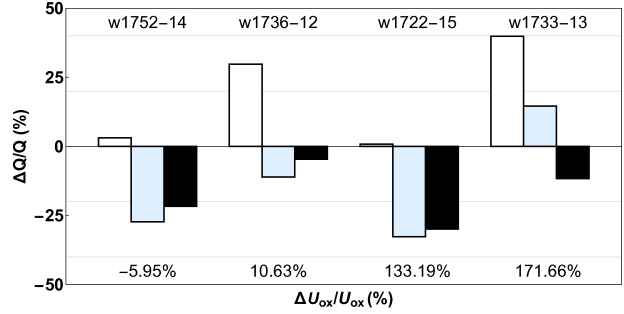}
\caption{ Sensitivity of the relative variations of the $Q$-factor, $\Delta Q/Q$, with respect to the relative variation of the undercut $\Delta U_{ox}/U_{ox}$ for the first three modes: (white) mode indexes (01);  (light-blue) mode indexes (11); (black) mode indexes (21).
 The relative variations are deduced from the values measured in the reference device w1726-14.}
\label{fig:IEEE_fig_17}
\end{figure}
%-----------------

\begin{table}
\centering
\caption{Ellipsometry measurement of the thickness of the TEOS oxide and the SiN and the 
average undercut after HF etching.}
\small
\begin{tabular}{cccc}
DEVICE & $\rm t_{ox}\pm 0.5 \%  $ & $\rm t_{SiN} \pm 0.5 \%  $ & $\rm u \pm 5 \% $\\ 
       & $[nm] $ & $[nm]$ & $[\mu m]$\\
\hline
\hline
w1722-15 & 298  & 108   & 5.48\\
w1726-14 & 298  & 100   & 2.35\\
w1733-13 & 302  & 109   & 6.37\\
w1736-12 & 302  & 109   & 2.60\\
w1752-14 & 298  & 100   & 2.21\\ 
\hline
\hline
\end{tabular}
\label{tab:IEEE_table_3}
\end{table}

%-----------------
\begin{table}
\centering
\caption{Experimental measurements of device parameters and mechanical quality factor at room temperature (300 K).}
\small
\begin{tabular}{ccccccc}
DEVICE & $\sigma_0$   & $ D $  &  $ \rm \nu_{mn} $  & Q & $ n_0 \, n_1 $ & $\rm f \times Q$\\
 & $[\rm MPa]$ & $[\rm mm]$ & $[\rm kHz]$ & $\rm [10^6]$ & &  $ \rm [THz]$ \\
\hline
\hline
    &    &     &      247.47 &  $\rm 8.5 $ &(01) & 2.09 \\
w1722-15 &  703.07  & 1.55   &  374.38 & $\rm 8.4 $ &(11) & 3.14 \\
	     &    &     &      568.29 & $\rm 7.8 $ &(21) & 4.42\\ 
\hline
      & &     &       233.98 & $\rm 8.4 $& (01) & 1.96\\
w1726-14	   & 712.22  & 1.65  &  372.96 & $\rm 12.5$& (11) & 4.65 \\
		   &  &     &       499.95 & $\rm 11.1$& (21)& 5.56  \\	        
\hline
    &   &     &    274.16& $\rm 11.7 $ &(01)& 3.22 \\
w1733-13  & 862.90   & 1.55   &   438.82 & $\rm 14.3 $ &(11)& 6.27\\
		  &   &     &    718.95 & $\rm  9.8 $& (21) & 7.06 \\
\hline
       &   &     &    275.91 & $\rm 10.9 $ &(01) & 3.00 \\
w1736-12	& 818.48   & 1.50  &   439.49 & $\rm 11.1 $& (11) & 4.87 \\
   		    &   &     &    589.91 & $\rm 10.6 $& (21)& 6.25\\
\hline
         &  &     &     241.82 & $\rm 8.7 $& (01) & 2.09\\
w1752-14	   &  671.33  & 1.55    & 384.88 & $\rm 9.1 $ &(11)& 3.49 \\
		   &  &     &     516.77 & $\rm 8.7 $& (21)& 4.50\\
\hline
\hline
\end{tabular}
\label{tab:IEEE_table_4}
\end{table}

The undercut  $\rm U_{ox}$ is roughly proportional to the etching time chosen for the release of the SiN membrane (Fig. \ref{fig:IEEE_fig5} step 8). Actually it is used to check if the SiN membrane is fully released from the SiO2 layer underneath: no undercut means no free-standing membrane. Starting form the reference process, corrisponding to device w1726-14, different etching time were tested to check possible effects on the quality factor. For each wafer the undercut, measured with Leica DM750M using a magnification of 50$\times$, is 
reported in Table \ref{tab:IEEE_table_3}. As shown in Fig. \ref{fig:IEEE_fig_17}, where no clear correlation can be seen, the undercut does not influence the $Q$-factor of the modes. Here the relative variations are evaluated from the reference device.

%-------------------
\section{Conclusion}
%-------------------
This work presents the  microfabrication process for the production of high-Q SiN membranes with an integrated on-chip 3D loss-shield. Within its operating bandwidth (0-5 MHz), the device preserves the  intrinsic Q-factor of the pretensioned round-shaped nanomembrane, regardless of the modal order or resonant frequency. In particular, $Q$-factors at room temperature are about $10^7$, achieving the minimum requirement for room-temperature quantum optomechanics $Q\times f > 6.2\,$THz for all modes with frequency higher than $600\,$kHz.

The device is obtained by MEMS bulk-micromachining via Deep-Reaction Ion Etching (DRIE), combined with a two-side processing on a thick SOI wafer. The microfabrication process is quite flexible and additional layers could be deposited over the SiN membrane before the DRIE steps, if required for a sensing application. Therefore, such oscillator is a promising candidate for quantum sensing applications in the context of the emerging field of Quantum Technologies.


\end{document}